\documentclass{ws-ijmpc}

\begin{document}

\markboth{Nuno Crokidakis}
{Noise and disorder: phase transitions and universality in a model of opinion formation}

\catchline{}{}{}{}{}

\title{Noise and disorder: phase transitions and universality in a model of opinion formation}

\author{Nuno Crokidakis}

\address{Instituto de F\'isica, Universidade Federal Fluminense \\
Niter\'oi/RJ, Brazil\\
nuno@if.uff.br}

\maketitle

\begin{history}
\received{Day Month Year}
\revised{Day Month Year}
\end{history}

\begin{abstract}
In this work we study a 3-state opinion formation model considering two distinct mechanisms, namely independence and conviction. Independence is introduced in the model as a noise, by means of a probability of occurrence $q$. On the other hand, conviction acts as a disorder in the system, and it is introduced by two discrete probability distributions. We analyze the effects of such two mechanisms on the phase transitions of the model, and we found that the critical exponents are universal over the order-disorder frontier, presenting the same universality class of the mean-field Ising model. In addition, for one of the probability distributions the transition may be eliminated for a wide range of the parameters.

\keywords{Opinion dynamics; computer simulations; phase transitions; universality}

\end{abstract}

\ccode{PACS Nos.: 05.10.-a, 05.70.Jk, 87.23.Ge, 89.75.Fb}

\section{Introduction}

\qquad In the last thirty years, the statistical physics techniques have been successfully applied in the description of social and economic phenomena. Among the studied problems we can cite opinion dynamics, language evolution, biological aging, dynamics of stock markets, earthquakes and many others \cite{galam_book,sen_book,pmco_book}. These subjects are interdisciplinary, and they are in general treated by means of computer simulations of agent-based models, which allow us to understand the emergence of collective phenomena in those systems.

Among typical problems of interest in social models, we highlight the dynamics of opinion formation \cite{galam,lalama,sznajd_indep1,sznajd_indep2,sznajd_indep3,nuno_pla,nuno_celia,lccc,sen,biswas_conf,deffuant2,brugna,xiong,nuno_pmco_jstat,nuno_bjp,biswas11,biswas,javarone}. Recently, the impact of nonconformity in opinion dynamics has been atracted attention of physicists, with many works published \cite{galam,lalama,sznajd_indep1,sznajd_indep2,sznajd_indep3,nuno_pla}. One class of nonconformist is the Anticonformist, that is similar to a conformist, since both take cognizance of the group norm. Thus, conformers agree with the norm, anticonformers disagree. Another class is formed by individuals presenting the Independent behavior, where the individual tends to resist to the groups' influence. As discussed in recent works \cite{sznajd_indep2,sznajd_indep3}, independence acts on an opinion model as a kind of stochastic driving that can lead the model to undergoes a phase transition, a similar effect produced by a social temperature \cite{lalama,sznajd_indep2,sznajd_indep3,nuno_pla}. On the other hand, conviction is another mechanism that has been considered recently in models of opinion formation \cite{nuno_celia,lccc,sen,biswas_conf,deffuant2,brugna,xiong,nuno_pmco_jstat,nuno_bjp}.

In this work we study the impact of independence and conviction on agents' behavior in a discrete kinetic opinion model. In particular, we introduce a probability $q$ of agents to make independent decisions, as well as conviction as a disorder in the system, that is considered in the model during the pairwise interactions. We perform computer simulations of an agent-based model, and the results show that the model undergoes phase transitions at critical points $q_{c}(w)$ that depend on a parameter $w$, related to the agents' convictions. The exponents of these transitions are universal, and the model is in the same universality class of the mean-field Ising model. Some of our results are complemented by analytical calculations.

This work is organized as follows. In Section 2 we present the microscopic rules that define the model and in Section 3 the numerical and analytical results are discussed. Finally, our conclusions are presented in Section 4.


\section{Model}

\qquad Our model is based on kinetic exchange opinion models (KEOM's) \cite{lccc,sen,biswas_conf,biswas11,biswas}. A population of $N$ agents is defined on a fully-connected graph, i.e., each agent can interact with all others, which characterizes a mean-field-like scheme. In addition, each agent $i$ carries one of three possible opinions (or states), namely $o_{i}=+1$, $-1$ or $0$. The following microscopic rules govern the dynamics:

\begin{enumerate}

\item An agent $i$ is randomly chosen;

\item With probability $q$, this agent will act independently. In this case, with equal probabilities ($1/3$) he/she chooses one of the possible opinions, $o_{i}=0$, $o_{i}=+1$ or $o_{i}=-1$;

\item On the other hand, with probability $1-q$ we choose another agent, say $j$, at random, in a way that $j$ will influence $i$. Thus, the opinion of the agent $i$ in the next time step $t+1$ will be updated according to
\begin{equation}\label{eq0}
o_{i}(t+1) = {\rm sgn}\left[c_{i}\,o_{i}(t) + o_{j}(t)  \right]\,,
\end{equation}
where the sign function is defined such that ${\rm sgn}(0)=0$.
\end{enumerate}

The model has two ingredients, namely noise and disorder. Noise is introduced by the independent behavior, represented by the probability $q$. On the other hand, the disorder is introduced in the system by the term $c_{i}$, that is characteristic of each agent $i$. In this case, $c_{i}$ is a stochastic variable that can be interpreted as the conviction of the agent $i$, and it can follow one of two possible probability distributions, namely
\begin{eqnarray} \label{eq1}
F_{1}(c_{i}) & = & w\,\delta(c_{i}-1) + (1-w)\,\delta(c_{i}) ~, \\ \label{eq1-1}
F_{2}(c_{i}) & = & w\,\delta(c_{i}-1) + (1-w)\,\delta(c_{i}+1) ~.
\end{eqnarray}

In other words we have a diluted ($+1,0$) and a bimodal ($\pm 1$) distributions of convictions. The interpretation of such values can be understood as follows. A given agent $i$ with conviction $c_{i}=+1$ is an individual $100\%$ aware of his/her opinion, that cannot be easily persuaded by another agent $j$. On the other hand, a value of conviction $c_{i}=-1$ represents an individual $i$ with a tendency to spontaneously change his/her opinion, since the first term of Eq. (\ref{eq0}) change from $o_{i}(t)$ to $-o_{i}(t)$. Finally, the value $c_{i}=0$ models an agent $i$ with no conviction about his/her opinion, i.e., an individual who can be easily persuaded during an interaction with a randomly chosen agent $j$.

Thus, the parameter $w$ quantifies the fraction of positive convictions, and the complementary fraction $1-w$ of convictions may assume the value $0$ or $-1$. In the case where the agent $i$ does not act independently, the change of his/her state occur according to a rule similar to the one proposed recently in the KEOM's \cite{nuno_pla,nuno_celia,biswas}. Notice, however, that in Ref. \refcite{nuno_pla} the population is homogeneous, i.e.,  all agents present the same conviction $c_{i}=1$ (no disorder). In this case, the model of Ref. \refcite{nuno_pla} undergoes a nonequilibrium phase transition at $q_{c}=1/4$. On the other hand, in the absence of independence and in the presence of the conviction term, the model does not present a phase transition \cite{nuno_celia}. Our objective in this work is to analyze the impact of the presence of the two mechanisms together, disorder and noise, in the phase transitions.

Thus, our Eq. (\ref{eq0}) represents the KEOM of Ref. \refcite{biswas} with no negative interactions, but with disorder (conviction) and noise (independence). In this case, for $q=w=0$ (no independence and no disorder) all stationary states will give us a population sharing consensus with $m=1$, where $m$ is the order parameter of the system, 
\begin{eqnarray} \label{eq2}
m = \left\langle \frac{1}{N}\left|\sum_{i=1}^{N}\,o_{i}\right|\right\rangle ~, 
\end{eqnarray}
and $\langle\, ...\, \rangle$ denotes a disorder or configurational average taken at steady states. The Eq. (\ref{eq2}) defines the ``magnetization per spin'' of the system. We will show by means of analytical and numerical results that the presence of the conviction term together with the independent behavior induces phase transitions in the KEOM in the absence of negative interactions. 

The three states considered in the model can be interpreted as follows. We have a population participating in a public debate with two distinct choices, for example a referendum with choices \textit{yes/no}. Thus, the opinions represent an agent in favor of \textit{yes} (opinion $+1$), in favor of \textit{no} (opinion $-1$), or indecision (opinion $0$). Notice that the undecided agents are not the same as independent ones. Indeed, an agent $i$ that decide to behave independently (with probability $q$) can make a decision to change or not his/her opinion (\textit{yes} to \textit{no}, for example) based on his/her own conviction, whatever is the his/her current state $o_{i}$ (decided or undecided). In other words, an interaction with another individual $j$ is not required. On the other hand, an undecided agent $i$ can change his/her opinion $o_{i}$ in two ways: due to an interaction with a decided agent $j$ (following the rule given by Eq. (\ref{eq0}), with probability $1-q$) or due to his/her own decision to do that (independently, with probability $q$).

The numerical procedure is as follows. At each time step we apply the above-mentioned rules $N$ times, where $N$ is the population size. As the model is a 3-state probabilistic cellular automaton \cite{biswas11}, we applied a parallel (synchronous) update scheme. In addition, we considered annealed random variables $c_{i}$, that are generated according to Eqs. (\ref{eq1}) or (\ref{eq1-1}) at each interaction with a pair of agents $i$ and $j$. This choice is due to test the analytical calculations presented in the Appendix, that are done for the annealed case. In addition, the initial configuration of the population in each simulation is fully disordered, i.e., we started all simulations with an equal fraction of each opinion ($1/3$ for each one).


\section{Results}   

\qquad Summarizing the microscopic rules of the previous section, the probability that an agent $i$ chooses a given opinion $+1$, $-1$ or $0$ independently of the opinions of the other agents is $q/3$. On the other hand, with probability $1-q$ we randomly choose another agent $j$ and apply the kinetic exchange rule given by Eq. (\ref{eq0}). For the analysis of the model, we have considered the order parameter $m$ defined by Eq. (\ref{eq2}), as well as the susceptibility $\chi$ and the Binder cumulant $U$ \cite{binder}, defined as
\begin{eqnarray} \label{eq3}
\chi & = &  N\,(\langle O^{2}\rangle - \langle O \rangle^{2}) \, ~ \\ \label{eq4}
U &  = &  1 - \frac{\langle O^{4}\rangle}{3\,\langle O^{2}\rangle^{2}} \,.
\end{eqnarray}


\subsection{Diluted distribution of convictions}

\begin{figure}[t]
\begin{center}
\vspace{3mm}
\includegraphics[width=0.55\textwidth,angle=0]{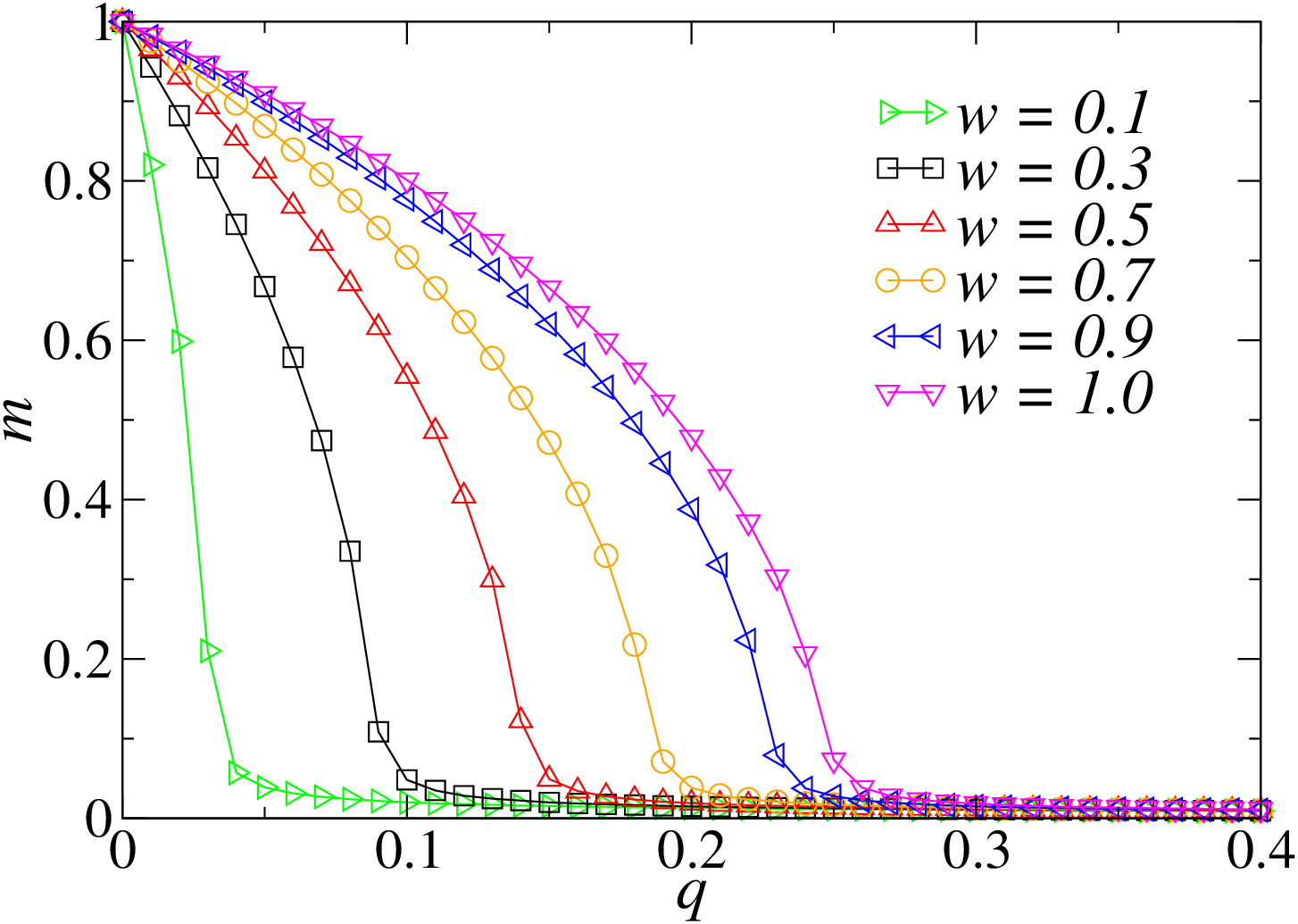}
\end{center}
\caption{(Color online) Order parameter $m$ as a function of the independence probability $q$ for typical values of $w$. The convictions follow the diluted distribution $F_{1}(c_{i})$ given by Eq. (\ref{eq1}). The population size is $N=10^{4}$ and results are averaged over $150$ samples.}
\label{fig1}
\end{figure}

\qquad In Fig. \ref{fig1} we exhibit the order parameter as a function of the independence probability $q$ for typical values of the conviction parameter $w$, for the diluted distribution $F_{1}(c_{i})$ given by Eq. (\ref{eq1}), for $N=10^{4}$. One can observe a typical behavior of a phase transition, that occurs for distinct points $q_{c}(w)$. Consensus states with $m=1$ occurs only for $q=0$, and there is a majority opinion in the population for any value of $0<q<q_{c}(w)$. For each value of $w$ we have an order-disorder transition, where in the ordered phase one of the extreme opinions ($+1$ or $-1$) is the majority opinion in the population, i.e., there is a decision in the public debate. On the other hand, in the disordered phase there is a coexistence of the three possible opinions, and the stationary fractions of the opinions are equal to $1/3$ (see the Appendix A.1). In this case, it means that there is no clear decision in the debate. 

\begin{figure}[t]
\begin{center}
\vspace{3mm}
\includegraphics[width=0.42\textwidth,angle=0]{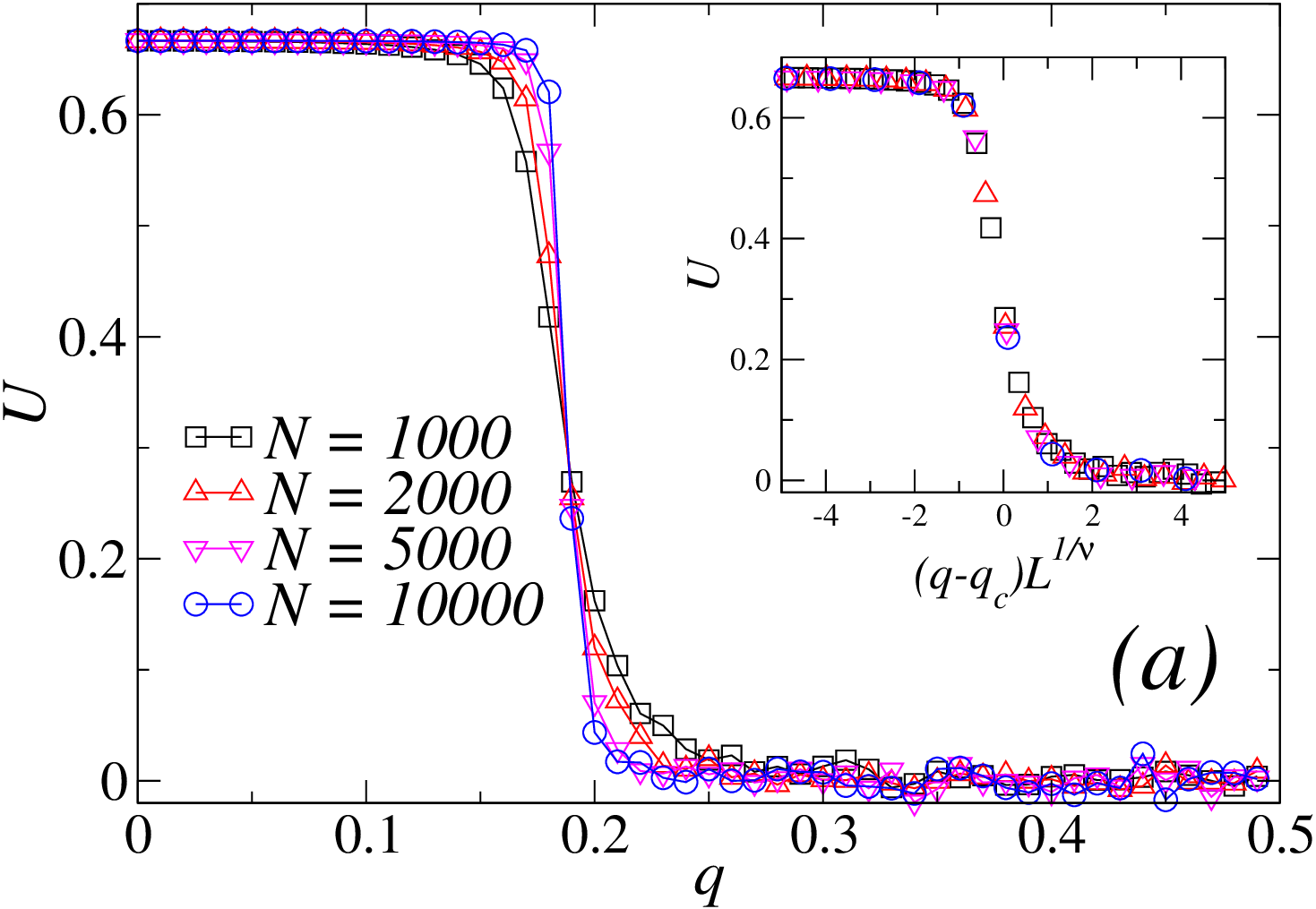}
\hspace{0.4cm}
\includegraphics[width=0.42\textwidth,angle=0]{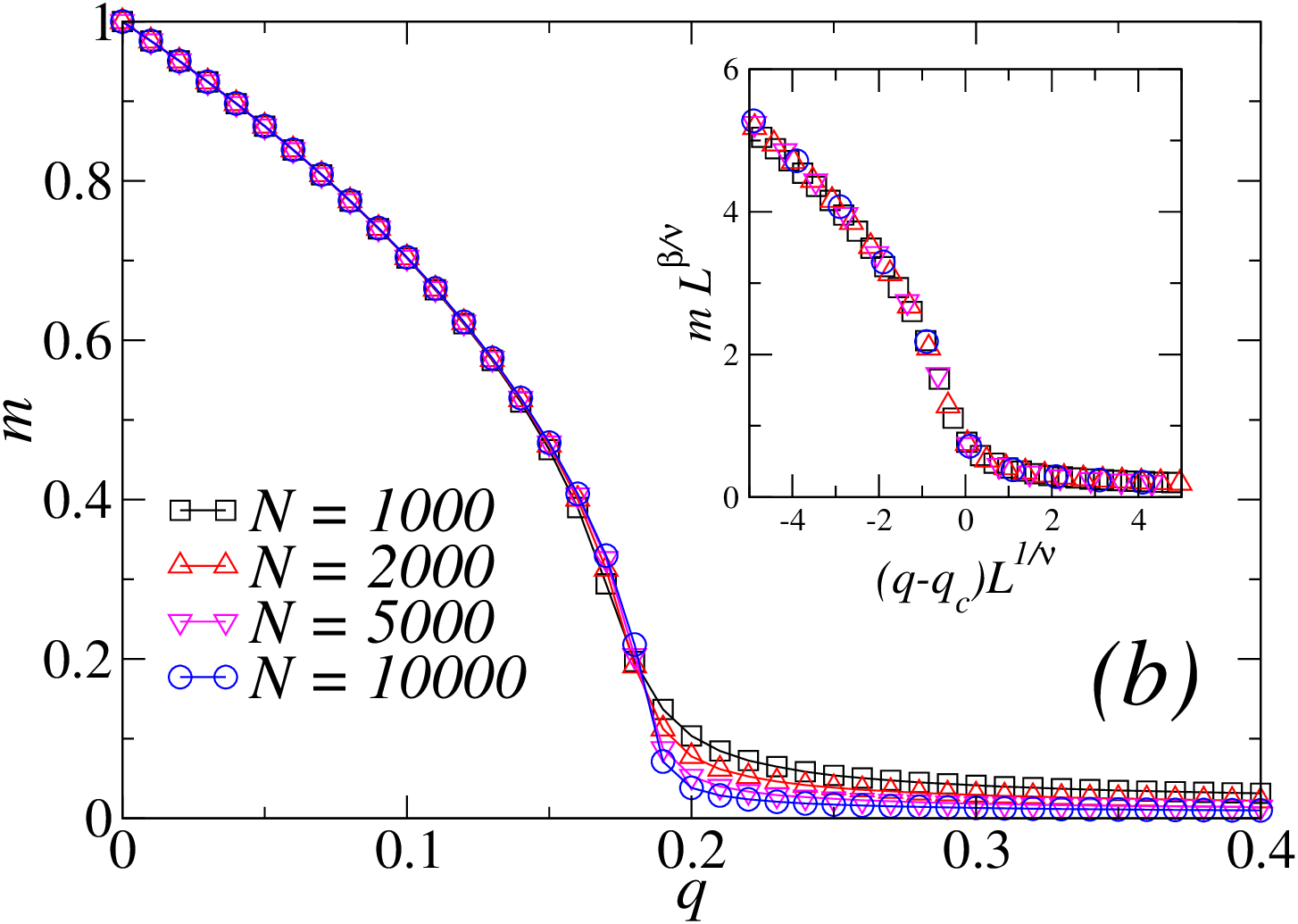}
\\
\vspace{0.8cm}
\includegraphics[width=0.42\textwidth,angle=0]{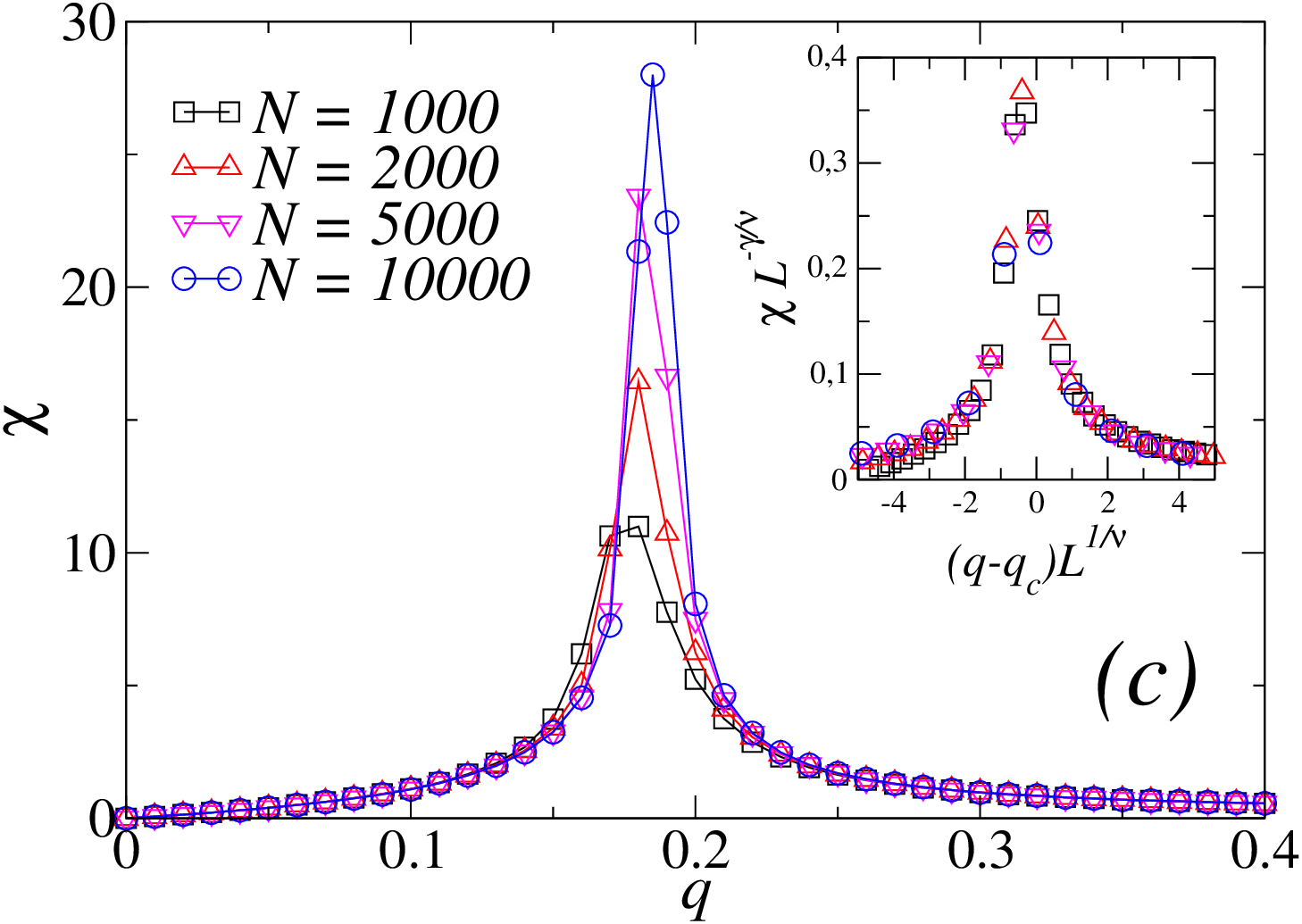}
\end{center}
\caption{(Color online) Binder cumulant $U$ ($a$), order parameter $m$ ($b$) and susceptibility $\chi$ ($c$) as functions of the independence probability $q$ for $w=0.7$ and distinct population sizes $N$. In the inset we exhibit the corresponding scaling plots. The estimated critical quantities are $q_{c}\approx 0.189$, $\beta\approx 0.5$, $\gamma\approx 1.0$ and $1/\nu\approx 0.5$. Results are averaged over $300$, $250$, $200$ and $150$ samples for $N=1000, 2000, 5000$ and $10000$, respectively.}
\label{fig2}
\end{figure}

In order to estimate the critical points, we look for the crossing of the Binder cumulant curves for the different sizes \cite{binder,salinas}, for each value of $w$. The corresponding critical exponents were estimated by a finite-size scaling (FSS) analysis, based on the standard scaling relations, 
\begin{eqnarray} \label{eq5}
m(N) & \sim & N^{-\beta/\nu} \\  \label{eq6}
\chi(N) & \sim & N^{\gamma/\nu} \\   \label{eq7}
U(N) & \sim & {\rm constant} \\   \label{eq8}
q_{c}(N) - q_{c} & \sim & N^{-1/\nu} ~,
\end{eqnarray}
that are valid in the vicinity of the phase transition. A typical example is shown in Fig. \ref{fig2}, where we exhibit results of simulations for $w=0.7$. For this case, our estimate is $q_{c}\approx 0.189$. For the critical exponents, we obtained $\beta\approx 0.5$, $\gamma\approx 1.0$ and $\nu\approx 2.0$. These expoents are universal, i.e., we found the same values for $\beta$, $\gamma$ and $\nu$ for all values of $w$. These results indicates that the order-disorder transition is universal, which is expected due to the mean-field character of the model.

Notice that the exponents $\beta$ and $\gamma$ are typical Ising mean-field exponents, which is not the case for $\nu$. This same discrepancy was observed in other discrete opinion models \cite{nuno_pla,nuno_celia,biswas}, and was associated with a superior critical dimension $d_{c}=4$, that leads to an effective exponent $\nu^{'}=1/2$, obtained from $\nu=d_{c}\,\nu^{'}=2$. In this case, one can say that our model is in the same universality class of the KEOM with two-agent interactions \cite{nuno_pla,nuno_celia,biswas11,biswas}, as well as in the mean-field Ising universality class.

\begin{figure}[t]
\begin{center}
\vspace{1.0cm}
\includegraphics[width=0.5\textwidth,angle=0]{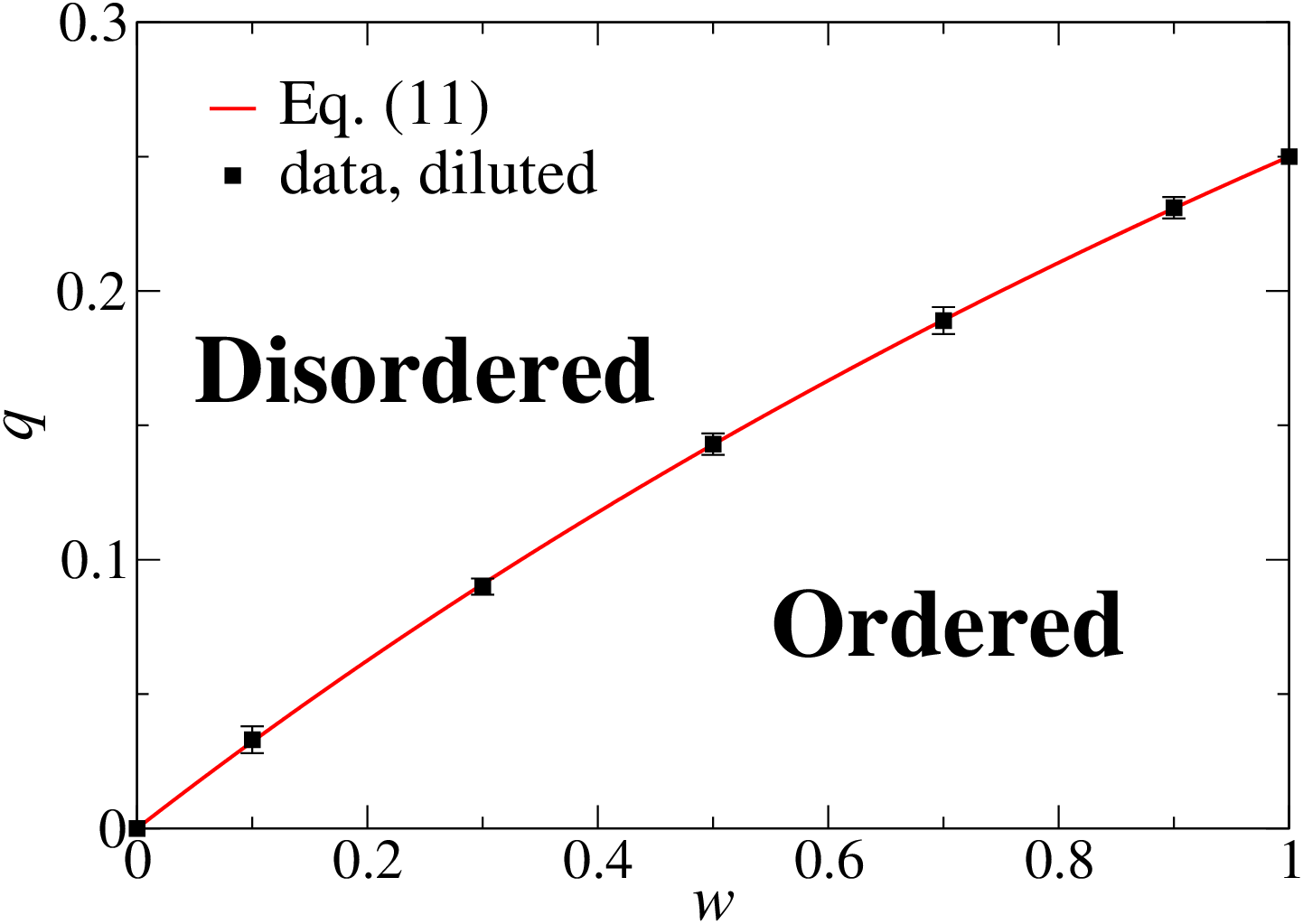}
\end{center}
\caption{(Color online) Phase diagram of the model in the plane $q$ versus $w$ for the diluted distribution of convictions, Eq. (\ref{eq1}). The diagram separates the ordered and the disordered phases. The symbols are the numerical estimates of the critical points $q_{c}(w)$, whereas the full line is the analytical prediction, Eq. (\ref{eq9}). The error bars were determined by the FSS analysis.}
\label{fig3}
\end{figure}

As above discussed, the numerical results suggest that critical points $q_{c}$ depend on $w$. This picture is confirmed by the analytical solution of the model, which give us (see Eq. (\ref{qc_sym}) of the Appendix A.1)
\begin{equation} \label{eq9}
q_{c}(w) = \frac{w}{w+3} \,.
\end{equation}
\noindent
Notice that the above solution give us $q_{c}(w=1) = 1/4$, which agrees with Ref. \refcite{nuno_pla}. Based on the numerical estimates of $q_{c}(w)$, we plot in Fig. \ref{fig3} the data together with the analytical result, Eq. (\ref{eq9}). One can see that the numerical results agree very well with the analytical prediction.

\subsection{Bimodal distribution of convictions}

\begin{figure}[t]
\begin{center}
\vspace{3mm}
\includegraphics[width=0.55\textwidth,angle=0]{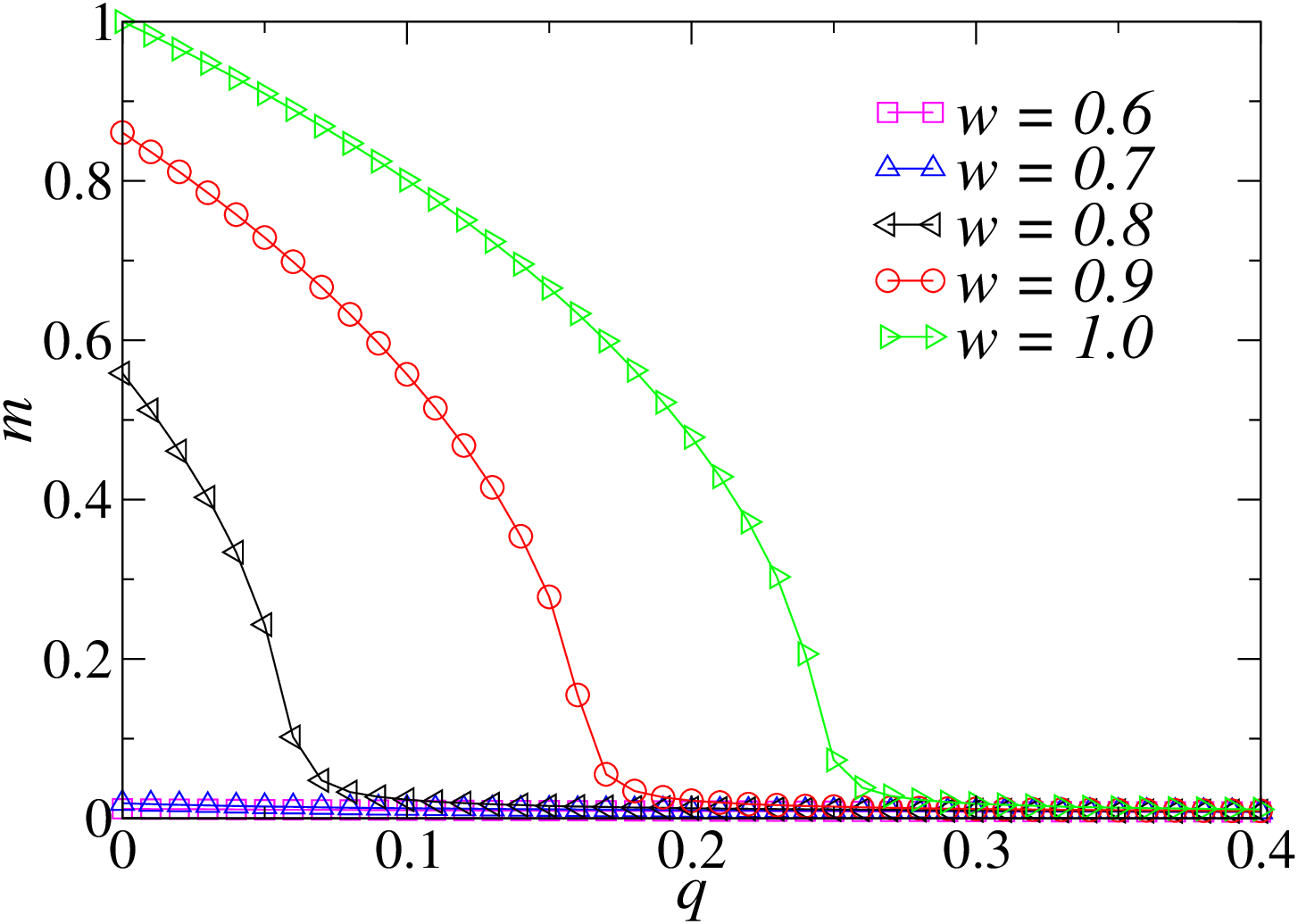}
\end{center}
\caption{(Color online) Order parameter $m$ as a function of the independence probability $q$ for typical values of $w$. The convictions follow the bimodal distribution $F_{2}(c_{i})$ given by Eq. (\ref{eq1-1}). The population size is $N=10^{4}$ and results are averaged over $150$ samples.}
\label{fig4}
\end{figure}

\begin{figure}[t]
\begin{center}
\vspace{1.0cm}
\includegraphics[width=0.5\textwidth,angle=0]{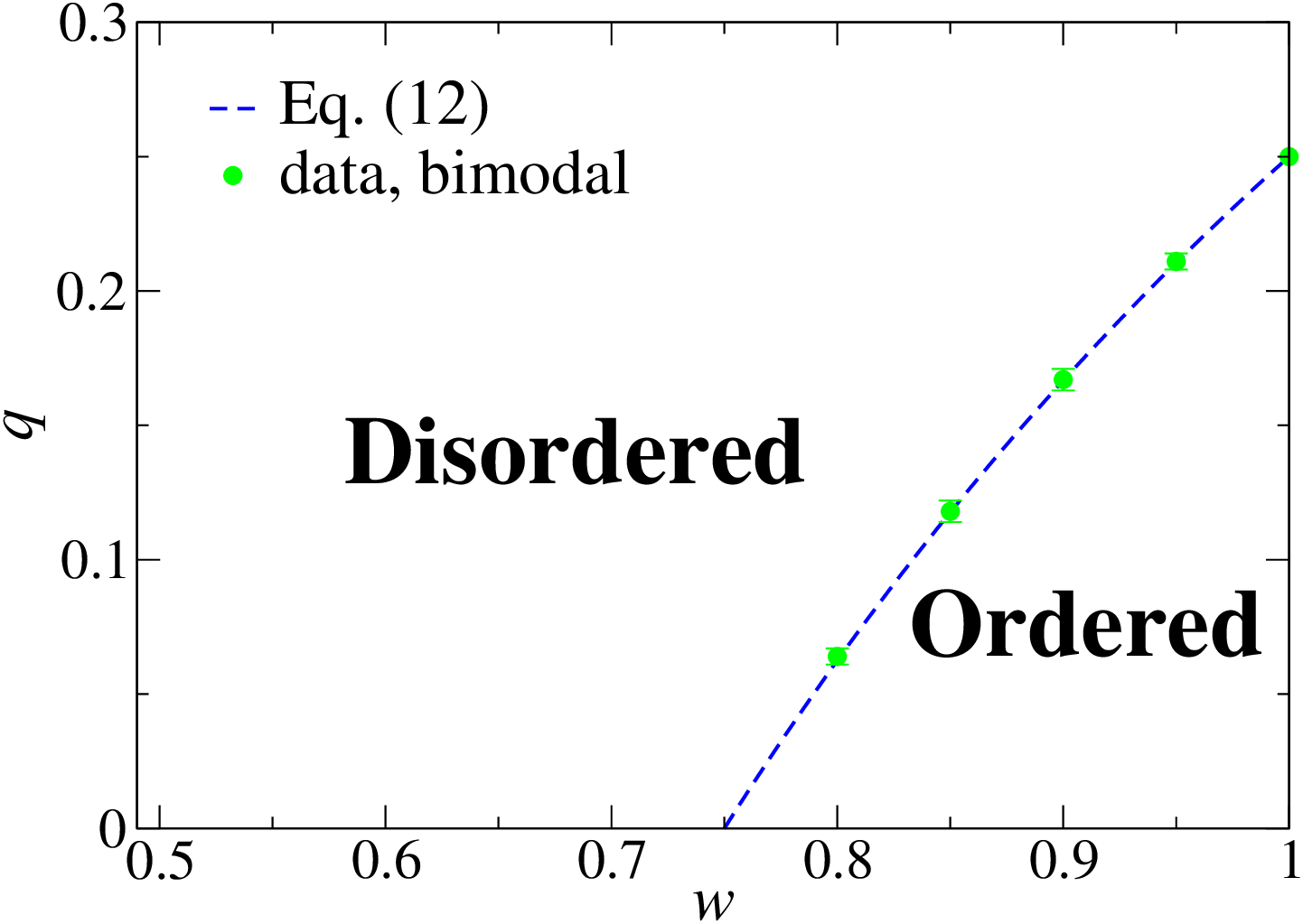}
\end{center}
\caption{(Color online) Phase diagram of the model in the plane $q$ versus $w$ for the bimodal distribution of convictions, Eq. (\ref{eq1-1}). The diagram separates the ordered and the disordered phases. The symbols are the numerical estimates of the critical points $q_{c}(w)$, whereas the full line is the analytical prediction, Eq. (\ref{eq10}). The error bars were determined by the FSS analysis. Notice that for $w<3/4$ there is no phase transition, in agreement with Eqs. (\ref{eq10}) and (\ref{eq11}).}
\label{fig5}
\end{figure}

\qquad In Fig. \ref{fig4} we exhibit the order parameter as a function of the independence probability $q$ for typical values of the conviction parameter $w$, for the bimodal distribution $F_{2}(c_{i})$ given by Eq. (\ref{eq1-1}), for $N=10^{4}$. As in the previous subsection, one can observe a typical behavior of a phase transition, that occurs for distinct points $q_{c}(w)$. However, one can see that the consensus $m=1$ occurs only for $q=0$ and $w=1$. For the other values of $w$, even for $q=0$ there is no consensus' states anymore. One can also see from Fig. \ref{fig4} that for sufficient small values of $w$ one can not observe a transition anymore, i.e., the system is disordered for all values of $q$. This picture is confirmed by the analytical solution of the model (see Eq. (\ref{qc_sym2}) of the Appendix A.2),
\begin{equation} \label{eq10}
q_{c}(w) = 1-\frac{3}{4w} \,.
\end{equation}
\noindent
As in the previous subsection, the above solution give us $q_{c}(w=1)=1/4$, which agrees with Ref. \refcite{nuno_pla}. However, different from the distribution $F_{1}(c_{i}$), for $F_{2}(c_{i}$) there is another transition point, given $q_{c}(w_{c})=0$, or 
\begin{equation} \label{eq11}
w_{c} = \frac{3}{4} \,.
\end{equation}
This result means that for $w<w_{c}=3/4$ there is no phase transition, independent on the value of $q$. In other words, the system is in a disordered phase for all values of the independence probability $q$. The numerical estimates of $q_{c}(w)$ are ploted in Fig. \ref{fig5} with the above analytical result, Eq. (\ref{eq10}). 

We found again the same critical exponents $\beta\approx 0.5$, $\gamma\approx 1.0$ and $\nu\approx 2.0$, over the order-disorder frontier, i.e., for $3/4<w\leq 1$. Thus, the phase transition is also universal in this case.

\section{Final Remarks}

\qquad In this work we studied a discrete model of opinion evolution and formation. For this purpose, we considered a fully-connected population where two distinct mechanisms act, namely conviction and independence, that play the role of disorder and noise, respectively, in the model.

We built the model considering a probability $q$ of an agent behave independently of the interactions, and with the complementary probability $1-q$ we consider pairwise interactions between two randomly chosen agents $i$ and $j$. In this case, the agent $j$ will try to persuade the agent $i$ through a kinetic exchange, and in this case the conviction $c_{i}$ of agent $i$ is considered. For this purpose, we considered two distinct probability distributions for the stochastic variables $c_{i}$, namely a diluted ($c_{i}=+1$ or  $0$) and a bimodal ($c_{i}=\pm 1$) distributions, in a way that a parameter $w$ denotes the fraction of positive ($c_{i}=+1$) convictions.

Computer simulations of the model indicate that the system undergoes order-disorder nonequilibrium phase transitions with universal exponents, and the model can be mapped in the mean-field Ising universality class. In the ordered phase one of the extreme opinions $+1$ or $-1$ is the majority in the population, indicating that there is a clear decision in the public debate. On the other hand, in the disordered phase the three opinions $+1, -1$ and $0$ coexist with equal fractions ($1/3$ for each one). For the bimodal distribution of convictions, the mentioned transition may be eliminated for $w<w_{c}=3/4$, i.e., if at least $25\%$ of the convictions in the population are negative ($c_{i}=-1$) we have a disordered state independently of the value of the parameter $q$. For both distributions the critical value $q_{c}(w)$ decreases for decreasing values of $w$, suggesting that in the presence of a fraction of agents that are not awared with their opinions, i.e., for $c_{i}=0$ or $-1$, it is more dificult to reach a decision if the population present individuals who behave as independents. Some of the results were confirmed by analytical calculations.

It would be interesting to extend the present model to the case where the agents' opinions are continuous. In this case, in addition to the analysis of the critical phenomena, it is interesting to analyze the distributions of opinions during the time evolution of the model and at the steady states. In this case, it can be discussed some usual characteristics of models with continuous states, namely the emergence and spread of extremist opinions, as well as moderate opinions, in both sides (positive and negative).


\section*{Acknowledgments}

The author acknowledges financial support from the Brazilian scientific funding agency CNPq.

\appendix

\section{Appendices}

\qquad Following Refs. \refcite{nuno_celia,biswas11,biswas}, we computed the critical values of the probability $q$ as a function of $w$ for the annealed case. We first obtained the matrix of transition probabilities whose elements $a_{i,j}$ give us the probability that a state suffers the change $i \to j$. Let us also define $f_{1}$, $f_{0}$ and $f_{-1}$, the stationary probabilities of each possible state, $+1$, $0$ and $-1$, respectively. In the steady state, the fluxes into and out from a given state must balance. For the  null state, one has
\begin{equation} \label{nullstate}
a_{1, 0}+a_{-1, 0}=a_{0,1}+a_{0,-1}   \,.
\end{equation}

Moreover, when the order parameter vanishes, we have $f_1=f_{-1}$. Finally, let us define $r(k)$, with $-2\le k \le 2$, the probability that the state change is $k$, that is, $r(k)=\sum_i a_{i,i+k}$. In the steady state, the average shift must vanish, namely, 
\begin{equation} \label{nullshift}
2\,r(2)-2\,r(-2) + r(1)-r(-1)=0 \,.
\end{equation}

In the following, we will consider separately the two probability distributions $F_{1}(c_{i})$ and $F_{2}(c_{i})$ of Eqs. (\ref{eq1}) and (\ref{eq1-1}), respectively.


\subsection{Diluted probability distribution $F_{1}(c_{i})$}

\qquad Considering the probability distribution $F_{1}$ for the convictions, Eq. (\ref{eq1}), the elements $a_{i,j}$ of the transition matrix are
 
\begin{eqnarray} \nonumber
a_{1, 1} &=& (q/3)\,f_{1} + (1-q)\,f_1^{2} + (1-q)\,w\,f_{1}\,f_{0} \\ \nonumber
a_{1, 0} &=& (q/3)\,f_{1} + (1-q)\,w\,f_{1}\,f_{-1} + (1-q)\,(1-w)\,f_{1}\,f_{0}     \\ \nonumber
a_{1, -1} &=& (q/3)\,f_{1} + (1-q)\,(1-w)\,f_{1}\,f_{-1} \\ \nonumber
a_{0, 1} &=& (q/3)\,f_{0} + (1-q)\,f_{0}\,f_{1}   \\ \nonumber
a_{0, 0} &=& (q/3)\,f_{0} + (1-q)\,f_{0}^{2}  \\ \nonumber
a_{0, -1} &=& (q/3)\,f_{0} + (1-q)\,f_{0}\,f_{-1}  \\ \nonumber
a_{-1, 1} &=& (q/3)\,f_{-1} + (1-q)\,(1-w)\,f_{1}\,f_{-1} \\ \nonumber
a_{-1, 0} &=& (q/3)\,f_{-1} + (1-q)\,w\,f_{1}\,f_{-1} + (1-q)\,(1-w)\,f_{0}\,f_{-1} \\ \nonumber
a_{-1, -1} &=& (q/3)\,f_{-1} + (1-q)\,f_{-1}^{2} + (1-q)\,w\,f_{0}\,f_{-1}  
\end{eqnarray}

In this case, the null state condition (\ref{nullstate}) give us two solutions in the disordered phase. The first one is $f_{0}=1/3$, which imples in $f_{1}=f_{-1}=1/3$ (disorder condition). The second solution is $f_{0}=1+[2q/3w(1-q)]$. The first solution satisfies all the physical requirements (normalization condition $f_{1} + f_{-1} + f_{0} = 1$, fractions $f_{i}<1$ for $i=-1,0,1$, etc), but the second solution does not. Indeed, the mentioned second solution gives us $f_{1}<0$ and $f_{-1}<0$ in the disordered phase. These results mean that the second solution is mathematically valid but it is physically unacceptable. Summarizing, the valid solution in the disordered phase is 
\begin{equation} \label{disorder_solution}
f_{1}=f_{-1}=f_{0}=1/3.
\end{equation}
On the other hand, the null average shift condition (\ref{nullshift}) gives us
\begin{equation} \label{null_solution}
f_{0} = \frac{q}{w\,(1-q)} \,,
\end{equation}
that is valid in the ordered phase. At the critical point both solutions (\ref{disorder_solution}) and (\ref{null_solution}) are valid, which results in
\begin{equation} \label{qc_sym}
q_{c}(w) = \frac{w}{w+3} \,.
\end{equation}
 This last expression gives us the order-disorder frontier in the $q$ versus $w$ phase diagram, exhibited in Fig. \ref{fig3}.


\subsection{Bimodal probability distribution $F_{2}(c_{i})$}

\qquad On the other hand, for the case of the probability distribution $F_{2}$ for the convictions, Eq. (\ref{eq1-1}), the elements $a_{i,j}$ of the transition matrix are
 
\begin{eqnarray} \nonumber
a_{1, 1} &=& (q/3)\,f_{1} + (1-q)\,w\,f_1^{2} + (1-q)\,w\,f_{1}\,f_{0} \\ \nonumber
a_{1, 0} &=& (q/3)\,f_{1} + (1-q)\,(1-w)\,f_{1}^{2} + (1-q)\,w\,f_{1}\,f_{-1}   \\ \nonumber
a_{1, -1} &=& (q/3)\,f_{1} + (1-q)\,(1-w)\,f_{1}\,f_{-1} + (1-q)\,(1-w)\,f_{1}\,f_{0} \\ \nonumber
a_{0, 1} &=& (q/3)\,f_{0} + (1-q)\,f_{0}\,f_{1}   \\ \nonumber
a_{0, 0} &=& (q/3)\,f_{0} + (1-q)\,f_{0}^{2}  \\ \nonumber
a_{0, -1} &=& (q/3)\,f_{0} + (1-q)\,f_{0}\,f_{-1}  \\ \nonumber
a_{-1, 1} &=& (q/3)\,f_{-1} + (1-q)\,(1-w)\,f_{1}\,f_{-1} + (1-q)\,(1-w)\,f_{0}\,f_{-1}\\ \nonumber
a_{-1, 0} &=& (q/3)\,f_{-1} + (1-q)\,w\,f_{1}\,f_{-1} + (1-q)\,(1-w)\,f_{-1}^{2} \\ \nonumber
a_{-1, -1} &=& (q/3)\,f_{-1} + (1-q)\,w\,f_{-1}^{2} + (1-q)\,w\,f_{0}\,f_{-1}  
\end{eqnarray}

In this case, the null state condition (\ref{nullstate}) give us two solutions in the disordered phase. The first one is $f_{0}=1/3$, which imples in $f_{1}=f_{-1}=1/3$ (disorder condition), and again the second solution, namely $f_{0}=1+[2q/3(1-q)]$ gives us $f_{1}<0$ and $f_{-1}<0$ in the disordered phase. These results mean that the second solution is mathematically valid but it is physically unacceptable. Summarizing, the valid solution in the disordered phase for this case is 
\begin{equation} \label{disorder_solution2}
f_{1}=f_{-1}=f_{0}=1/3.
\end{equation}
On the other hand, the null average shift condition (\ref{nullshift}) gives us
\begin{equation} \label{null_solution2}
f_{0} = \frac{1-w\,(1-q)}{w\,(1-q)} \,,
\end{equation}
that is valid in the ordered phase. At the critical point both solutions (\ref{disorder_solution2}) and (\ref{null_solution2}) are valid, which results in
\begin{equation} \label{qc_sym2}
q_{c}(w) = 1-\frac{3}{4w} \,.
\end{equation}
 This last expression gives us the order-disorder frontier in the $q$ versus $w$ phase diagram, exhibited in Fig. \ref{fig5}. Notice that there is another transition point given by $q_{c}(w_{c})=0$, or 
\begin{equation} \label{wc}
w_{c} = \frac{3}{4} \,.
\end{equation}

\end{document}